\documentclass[11pt]{article}
\usepackage[dvips]{color}
\usepackage{epsfig}
\usepackage{amsmath}
\usepackage{graphicx}
\usepackage{tabularx}
\date{}
\begin{document}

\title{{\bf A new holographic dark energy model in Brans-Dicke theory with logarithmic scalar field}}
\author{Ehsan Sadri\,\,
and Babak Vakili\thanks{b.vakili@iauctb.ac.ir}
\\\\
{\small {\it Department of Physics, Central Tehran Branch, Islamic
Azad University, Tehran, Iran}}}

\maketitle
\begin{abstract}
We study a holographic dark energy model in the framework of
Brans-Dicke (BD) theory with taking into account the interaction
between dark matter and holographic dark energy. We use the recent
observational data sets, namely SN Ia compressed Joint
Light-Analysis(cJLA) compilation, Baryon Acoustic Oscillations (BAO)
from BOSS DR12 and the Cosmic Microwave Background (CMB) of Planck
2015. After calculating the evolution of the equation of state as
well as the deceleration parameters, we find that with a logarithmic
form for the BD scalar field the phantom crossing can be achieved in
the late time of cosmic evolution. Unlike the conventional theory of
holographic dark energy in standard cosmology ($\omega_D=0$), our
model results a late time accelerated expansion. It is also shown
that the cosmic coincidence problem may be resolved in the proposed
model. We execute the statefinder and Om diagnostic tools and
demonstrate that interaction term does not play a significant role.
Based on the observational data sets used in this paper it seems
that the best value with $1\sigma$ and $2\sigma$ confidence interval
are $\Omega_m=0.268^{+0.008~+0.010}_{-0.007~-0.009}$, $
\alpha=3.361^{+0.332~+0.483}_{-0.401~-0.522}$,
$\beta=5.560^{+0.541~+0.780}_{-0.510~-0.729}$,
$c=0.777^{+0.023~+0.029}_{-0.017~-0.023}$ and $b^2 =0.045$,
according to which we find that the proposed model in the presence
of interaction is compatible with the recent observational data.
\vspace{5mm}\newline Keywords: Holographic dark energy; Brans-Dicke
theory; Coincidence problem\vspace{5mm}\newline PACS numbers:
95.36.+x, 95.35.+d, 98.80.-k

\end{abstract}

\section{Introduction}
Nowadays it is strongly believed that some exotic matter with
negative pressure, usually called dark energy (DE), is responsible
for the accelerated expansion of our universe. Although so far lots
of models have been proposed for dark energy, but it still remains
one of the open issues in cosmology \cite{1}. Among these models,
the holographic dark energy (HDE) has attracted much attention in
recent years \cite{2}. This proposal is based on the holographic
principle according to which all of the information saved in a
certain region of space can be extracted from its boundary area and
are limited by an IR cutoff \cite{3}. It can be shown that the
energy density of HDE is related to the cutoff, $L$, as
$\rho_D=3c^2M_p^2/L^2$ \cite{4}. Recalling that the black hole
entropy (and thus its information) is also related to its horizon
area (say $L^2$), it seems that the models of HDE are like the black
hole physics and modifications of black hole laws due to for
instance, modified and quantum gravity will affect the energy
density of HDE and so new cosmological consequences may arise from
it.

In this paper, we are going to study a HDE model in the context of
BD theory of gravity and by using of the combination of recent
observational data sets (SN Ia + BAO + CMB) we fix the relevant free
parameters. To do this, we will choose a logarithmic form for the BD
scalar field \cite{5} which its aim is to check better agreement
with observations and for the energy density of HDE we will take the
form which have been proposed in \cite{6}. In the presented model,
we deal with the recent accelerated expansion and some other
problems associated with the DE models like cosmic coincidence
problem in a flat FRW geometry. We also investigate the differences
between interacting and non-interacting modes and their relation to
resolve the coincidence problem.

\section{The model}
Let us start with the BD gravity theory characterized with the
action

\begin{equation}\label{A}
S=\int d^4x{\sqrt{-g}\Big[\frac{1}{2}\big(-\phi
R+\frac{w}{\phi}g^{\mu\nu}\partial_{\mu}\phi\partial_{\nu}\phi\big)+L_m\Big]},
\end{equation} where $R$ is the Ricci scalar, $w$ is the BD parameter, $\phi$ is the BD scalar field and $L_m$ denotes the matter
Lagrangian density. We assume a homogeneous and isotropic FRW
universe which is described by the line element

\begin{equation}\label{B}
ds^2=-dt^2+a^2(t)\left(\frac{dr^2}{1-kr^2}+r^2d\Omega ^2\right),
\end{equation}
where $k=0,\pm 1$ and $a(t)$ are the curvature index and scale
factor respectively. For a flat universe ($k=0$) the above action
leads the following equations of motion

\begin{equation}\label{C}
\frac{3}{4w}\phi^2
H^2-\frac{\dot{\phi}^2}{2}+\frac{3}{2w}H\dot{\phi}\phi=\rho_m+\rho_D,
\end{equation}

\begin{equation}\label{D}
\frac{-\phi^2}{4w}\left(\frac{2\ddot{a}}{a}+
H^2\right)-\frac{1}{w}H\dot{\phi}\phi-\frac{1}{2w}\ddot{\phi}\phi-\frac{\dot{\phi}^2}{2}\left(1+\frac{1}{w}\right)=p_D,
\end{equation}

\begin{equation}\label{E}
\ddot{\phi}+3H\dot{\phi}-\frac{3}{2w}\left(\frac{\ddot{a}}{a}+H^2\right)\phi=0,
\end{equation}where $\rho_D$ and $\rho_m$ are the energy densities of
DE and dark matter (DM) respectively, $p_D$ is the pressure of DE
and $H=\dot{a}/a$ is the Hubble function. As in \cite{5}, we assume
a logarithmic relation between BD scalar field and scale factor
which is claimed that it may avoid a constant result for the
deceleration parameter:

\begin{equation}\label{F}
\phi=\phi_0 \ln(\alpha+\beta a),
\end{equation}
where $\phi_0$, $\alpha>1$ and $\beta>0$ are some constants. In
\cite{6} a new model of HDE is proposed which is based on the DGP
braneworld theory. In such a extra dimensional cosmology the
four-dimensional FRW universe is embedded as a brane in a
five-dimensional flat (Minkowski) bulk. The motivations to add the
DE to this model is studied in \cite{7} and by adding DE to DGP in
\cite{6} a new modified model of HDE is resulted according to which
we can write the energy density of HDE as

\begin{equation}\label{G}
\rho_D=\frac{3c^2\phi^2}{4w L^2}(1-\frac{\epsilon L}{3 r_c}),
\end{equation}where $r_c=\frac{M^2_p}{2M^3_5}=\frac{G_5}{2G_4}$ stands for the
crossover length scale, $\epsilon=\pm1$ corresponds to the two
branches (self-accelerated and normal) of solution \cite{8} and
$L=H^{-1}$ is the Hubble horizon as the system's IR cutoff. The
$\epsilon=+1$ is related to the self-accelerating solution in which
the universe may enter an accelerating phase in the late time
without additional dark energy component. When $L\ll 3r_c$, above
equation reduces to usual HDE density. In what follows, by this
choice for the system's IR cutoff and fitting the free parameters by
use of the latest observational data, we study the evolution of DE
density, equation of state (EoS), deceleration parameters, ratio of
DE density to density of total matter of the universe and apply
diagnostic tools to find characteristics of the model.

\section{Interacting model}
\begin{figure}[t]
\begin{center}
\includegraphics[width=0.45\textwidth]{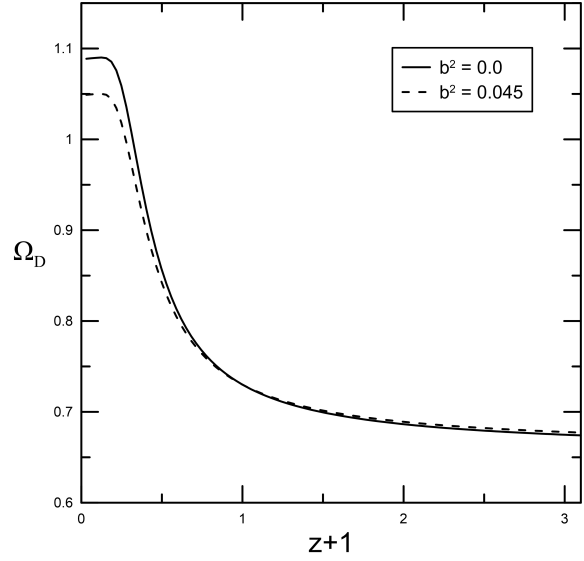}
\caption{Density parameter ($\Omega_D$) versus redshift.}
\label{fig1}
\end{center}
\end{figure}

In the presence of non-gravitational interaction between DE and DM
the continuity equations can be written as

\begin{equation}\label{S}
\dot{\rho}_m+3H\rho_m=Q,
\end{equation}

\begin{equation}\label{T}
\dot{\rho}_D+3H(1+\omega_D)\rho_D=-Q.
\end{equation}
where $\omega_D=\frac{p_D}{\rho_D}$ is the EoS parameter of DE. We define the dimensionless density parameters as

\begin{equation}\label{J}
\Omega_m=\frac{4w\rho_m}{3\phi^2H^2},
\end{equation}

\begin{equation}\label{K}
\Omega_D=c^2(1-\frac{\epsilon}{3Hr_c}),
\end{equation}
where $r_c=\frac{1}{4H^2\Omega_{r_c}}$. With $L=H^{-1}$ and using
(\ref{F}) the time derivative of equation (\ref{G}) results

\begin{equation}\label{L}
\dot{\rho}_D=\rho_D\left(\frac{\dot{\phi}}{\phi}+2\frac{\dot{H}}{H}\right)+2c^2\epsilon
\phi \sqrt{\Omega_{r_c}}\dot{H}{H}.
\end{equation}
The behavior of interaction with different $Q$-terms have been
studied in \cite{10}.\footnote{One of the most common choices for
$Q$-term is $3H(b^2_1\rho_D+b^2_2 \rho_m)$, in which $b^2_{1,2}$ are
the coupling constants, their values should be fixed by observation.
However, it is more appropriate to use a single coupling constant
such as: $Q_1=3b^2H\rho_D$, $Q_2=3b^2H\rho_m$ and
$Q_3=3b^2H(\rho_D+\rho_m)$.} In this work we take a simple
interaction term as $Q=3b^2H\rho_D$, by use of which and with the
help of equations (\ref{F}), (\ref{G}), (\ref{J}), (\ref{K}) and
(\ref{S}) we obtain

\begin{equation}\label{U}
\frac{\dot{H}}{H^2}=\frac{\frac{\beta a}{(\alpha+\beta
a)\ln(\alpha+\beta a)}\Omega_D-\frac{\beta^3a^3}{(\alpha+\beta
a)^3(\ln(\alpha+\beta a))^2}+\frac{\beta^2a^2}{(\alpha+\beta
a)^2\ln(\alpha+\beta a)}-2\frac{\beta a}{(\alpha+\beta
a)\ln(\alpha+\beta
a)}+\frac{3}{2}(\Omega_D(b^2+1)-1)}{1-\frac{\beta^2a^2}{(\alpha+\beta
a)^2(\ln(\alpha+\beta a))^2}+2\frac{\beta a}{(\alpha+\beta
a)\ln(\alpha+\beta a)}-\frac{1}{2}(c^2+\Omega_D)}.
\end{equation}
Taking time derivative of equation (\ref{K}) we obtain

\begin{equation}\label{N}
\dot{\Omega}_D=\frac{2}{3}\epsilon
c^2\sqrt{\Omega_{r_c}}\frac{\dot{H}}{H}.
\end{equation}
In order to see how the density parameter of HDE evolves, we define
$\Omega'_D= \frac{\dot{\Omega}_D}{H}$, where the prime denotes
derivative with respect to $x=\ln a$. Then from equation (\ref{U})
and (\ref{N}) we have
\begin{equation}\label{V}
\Omega'_D=(c^2-\Omega_D)\frac{\frac{\beta a}{(\alpha+\beta
a)\ln(\alpha+\beta a)}(\Omega_D-2)-\frac{\beta^3a^3}{(\alpha+\beta
a)^3(\ln(\alpha+\beta a))^2}+\frac{\beta^2a^2}{(\alpha+\beta
a)^2\ln(\alpha+\beta
a)}+\frac{3}{2}(\Omega_D(b^2+1)-1)}{1-\frac{\beta^2a^2}{(\alpha+\beta
a)^2(\ln(\alpha+\beta a))^2}+2\frac{\beta a}{(\alpha+\beta
a)\ln(\alpha+\beta a)}-\frac{1}{2}(c^2+\Omega_D)},
\end{equation}where again the prime denotes derivative with respect to $x=\ln a$.
The dimensionless HDE density parameter $\Omega_D$, as a function of
$1+z=a^{-1}$ is plotted in figure \ref{fig1}. As this figure shows,
at the early times ($z\rightarrow \infty$), density parameter tends
to zero and at the late time ($z \rightarrow -1$) it reaches to 1,
that is a DE dominated era and the coupling constant does not make a
major change to this behavior.

\begin{figure}[ht!]
\begin{tabular}{ccc}
\includegraphics[width=0.45\textwidth]{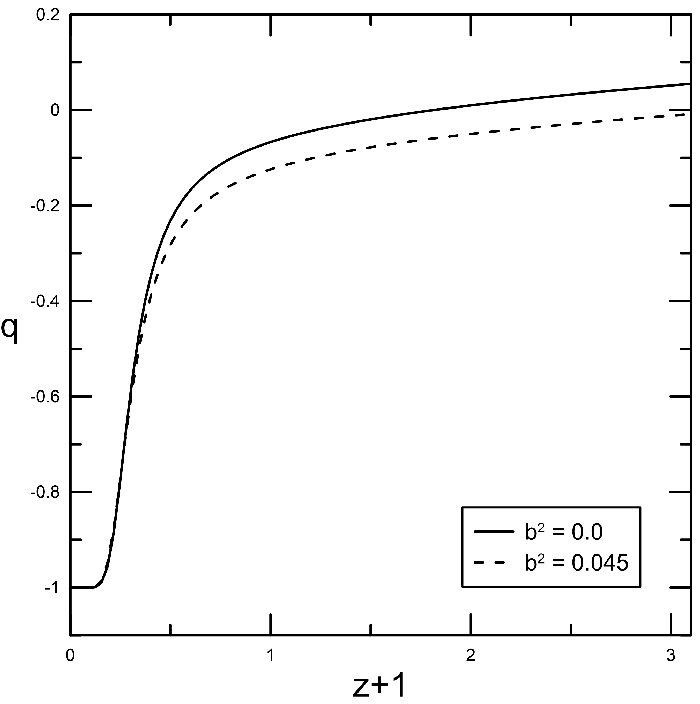}\hspace{2mm}
\includegraphics[width=0.45\textwidth]{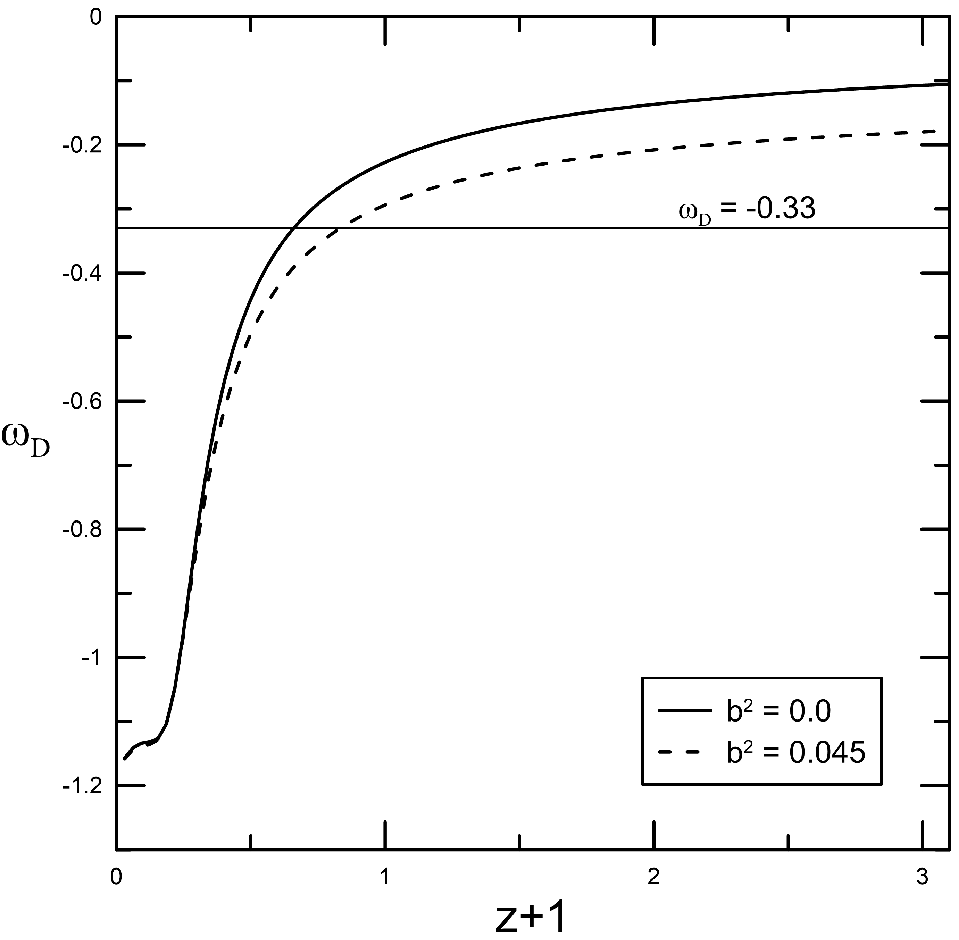}\end{tabular}
\caption{EoS and DP versus redshift.} \label{fig2}
\end{figure}

Now, let us compute the EoS parameter for DE. This may be done by
combining the equations (\ref{L}), (\ref{T}) and (\ref{U}) to get

\begin{equation}\label{W}
\begin{split}
\omega_D=-1-b^2-\frac{2\beta a}{3(\alpha+\beta a)\ln(\alpha+\beta
a)}~~~~~~~~~~~~~~~~~~~~~~~~~~~~~~~~~~~~~~~~~~~~~~~~~~~~~~~~~~~~~~~~~~~~~~~~~~~~~~~~~~~~~~~~~
\\ ~~~~  -(\frac{\Omega_D+c^2}{3\Omega_D})\frac{\frac{\beta
a}{(\alpha+\beta a)\ln(\alpha+\beta
a)}(\Omega_D-2)-\frac{\beta^3a^3}{(\alpha+\beta
a)^3(\ln(\alpha+\beta a))^2}+\frac{\beta^2a^2}{(\alpha+\beta
a)^2\ln(\alpha+\beta
a)}+\frac{3}{2}(\Omega_D(b^2+1)-1)}{1-\frac{\beta^2a^2}{(\alpha+\beta
a)^2(\ln(\alpha+\beta a))^2}+2\frac{\beta a}{(\alpha+\beta
a)\ln(\alpha+\beta a)}-\frac{1}{2}(c^2+\Omega_D)}.
\end{split}
\end{equation}
Another important parameter by means of which we can distinguish
accelerated or decelerated expansion is the deceleration parameter
(DP) with definition
\begin{equation}\label{Q}
q=-\frac{a\ddot{a}}{\dot{a}^2}=-1-\frac{\dot{H}}{H^2}.
\end{equation}
With the help of equations (\ref{U}) this parameter can be
calculated independently of the EoS parameter as

\begin{equation}\label{R}
q=-1-\frac{\frac{\beta a}{(\alpha+\beta a)\ln(\alpha+\beta
a)}\Omega_D-\frac{\beta^3a^3}{(\alpha+\beta a)^3(\ln(\alpha+\beta
a))^2}+\frac{\beta^2a^2}{(\alpha+\beta a)^2\ln(\alpha+\beta
a)}-2\frac{\beta a}{(\alpha+\beta a)\ln(\alpha+\beta
a)}+\frac{3}{2}(\Omega_D(b^2+1)-1)}{1-\frac{\beta^2a^2}{(\alpha+\beta
a)^2(\ln(\alpha+\beta a))^2}+2\frac{\beta a}{(\alpha+\beta
a)\ln(\alpha+\beta a)}-\frac{1}{2}(c^2+\Omega_D)}.
\end{equation}
The other form of the DP in terms of the EoS parameter comes from
equations (\ref{D})-(\ref{F}):
\begin{equation}\label{RR}
q=\frac{3\omega_D\Omega_D-\frac{2\beta^2a^2}{(\alpha+\beta
a)^2\ln(\alpha+\beta a)}+\frac{2\beta a} {(\alpha+\beta
a)\ln(\alpha+\beta
a)}+\frac{\beta^2a^2(1+\frac{1}{w})}{w(\alpha+\beta
a)^2\ln(\alpha+\beta a)^2}+1}{2+\frac{\beta a}{(\alpha+\beta
a)\ln(\alpha+\beta a)}}.
\end{equation}
In figures \ref{fig2} we have plotted the EoS and DP which both,
with high accuracy, show a transition from a deceleration phase to
an accelerating universe. We see that the sign of $\omega_D$ depends
on the sign of logarithm terms. In the early times when
$(a\rightarrow 0)$ all logarithmic terms tend to zero and finally we
get a function that  by choosing suitable parameters can reach the
phantom era $\omega_D<-1$ without necessity of interaction between
DE and DM. Also, in present time and by consideration of the fact
that for $r_c\gg 1$ we have $\Omega_{r_c}\rightarrow 0$ (and then
$\Omega_D=c^2$), we reach $\omega_D=0$, that is nothing but the
standard cosmology. There is no acceleration with Hubble horizon as
IR cutoff. Indeed, it is shown that EoS in non-interacting model
with Hubble horizon as IR cutoff cannot be able to define cosmic
acceleration \cite{9}. Despite of this fact, equation (\ref{W})
stemmed from DGP braneworld with logarithmic scalar field in BD
theory by using Hubble horizon demonstrates accelerated cosmic
expansion.

In late time the logarithmic terms vanish and approximately we have
$q\approx \frac{1+3\omega_D\Omega_D}{2}$. On the other hand, as is
clear from figures \ref{fig1} and \ref{fig2}, $\omega_D \rightarrow
-1$ in this limit and so $q\approx \frac{1-3\Omega_D}{2}$. Now, let
us to examine $\Omega_D$. For the late time $\Omega_D>1$, we have
calculated approximately ($\Omega_D=1.08$), then for DP we must have
($q=-1.31$) which the phantom realm is successfully recovered. For
all values $\Omega_D>1/3$, we have an accelerating expansion.
Therefore, one can infer that in the matter dominated universe
(early times) the expansion is decelerated and for DE dominated
universe (late times) we have accelerated expansion.

\section{Coincidence problem}
\begin{figure}[t]
\begin{center}
\includegraphics[width=0.45\textwidth]{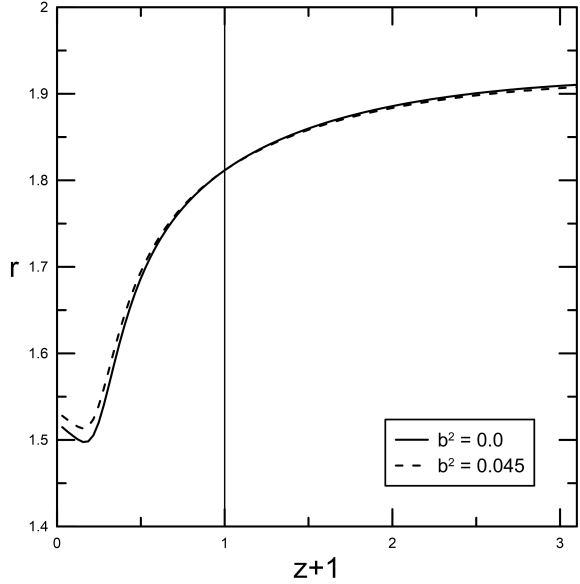}
\caption{Variation of $r$ in terms of the redshift.} \label{fig6}
\end{center}
\end{figure}

In this section we will deal with the cosmic coincidence problem
\cite{10-1}, in the framework of the presented model. Following the
formalism proposed in \cite{11} we determine the coupling value
between dark energy and dark matter. Let us define

\begin{equation}\label{X}
\zeta_{HDE}=1+\omega=\frac{\rho_{HDE}+p_{HDE}}{\rho_{HDE}},
\end{equation}

\begin{equation}\label{Y}
\zeta_{m}=\frac{\rho_{m}+p_{m}}{\rho_{m}},
\end{equation}
where $\zeta_m=1$ when $p_m=0$. To find how the density ratio
$r=\rho_m/\rho_D$ evolves with redshift, we have

\begin{equation}\label{Z}
\dot{r}=\frac{dr}{dt}=\frac{\rho_m}{\rho_{HDE}}\Big[\frac{\dot{\rho}_m}{\rho_m}-\frac{\dot{\rho}_{HDE}}{\rho_{HDE}}\Big].
\end{equation}With the use of (\ref{S}) and (\ref{T}) we have

\begin{equation}\label{AB}
\dot{r}=r\Big[\frac{Q}{\rho_m}+\frac{Q}{\rho_{NHDE}}+3H(\zeta_{HDE}-1)\Big],
\end{equation}which with $Q=3b^2(\rho_m+\rho_D)$, takes the form

\begin{equation}\label{AC}
\dot{r}=3H[r(\zeta_{HDE}-1)+(1+r)^2b^2].
\end{equation}Thus, the equation $\dot{r}=0$ has the solutions

\begin{equation}\label{AD}
r_{\pm}=\frac{1-\zeta_{HDE}}{2b^2}-1\pm
\Big[\Big(\frac{1-\zeta_{HDE}}{2b^2}-1\Big)^2-1\Big]^{\frac{1}{2}}.
\end{equation}
Since the negative $b^2$s violate the second law of thermodynamics
\cite{12}, we assume $b^2>0$. Indeed, observational constraints on
the coupling constant have shown that the coupling parameter is a
small positive value\cite{13}. Under this condition the expression
in the above square root will be positive if $\zeta\leq 1-4b^2$ or
$\zeta\geq1$. To address the cosmic coincidence problem \cite{14},
we rewrite the density ration as

\begin{equation}\label{AD}
r=-1+\frac{1}{\Omega_D}+\frac{2\beta a}{\Omega_D(\alpha+\beta
a)\ln(\alpha+\beta a)}-\frac{2}{3}\frac{\beta^2 a^2\omega }{\Omega_D
(\alpha+\beta a)^2 (\ln(\alpha+\beta a))^2},
\end{equation}in which we have used the relations (\ref{C}),
(\ref{F}), (\ref{G}) and (\ref{L}). This equation shows $r$ depends
on the scale factor, which in turn is a function of time. By
ignoring the last two terms in the early times of cosmic evolution
this equation reduces to $r=-1+\frac{1}{\Omega_D}$. However, as is
shown in \cite{14A} the rate of time variation of $r$ is very slow
compared to the scale factor. Therefore, with a good approximation
we may write

\begin{equation}\label{AE}
r_0\approx -1+\frac{1}{\Omega_D},
\end{equation}
where by $0$ we mean the value at the present time. Since according
to the observational data \cite{15}, the value of the parameter $c$
is restricted: $0.5<c<1$ , the value of $\Omega_D$ varies in the
interval $[0,1]$ and so we have $r_0\sim \mathcal{O}(1)$, see figure
\ref{fig6}. This analysis shows that although the value of the
parameter $c$ is important when one is going to determine the time
varying form of the value of $r$, it does not play a major role in
resolving the cosmic coincidence problem. The fact that $r_0\sim
\mathcal{O}(1)$ in the early and late times shows that the cosmic
coincidence is no longer seems to be a problem in this model.

\section{Statefinder diagnosis pair}
Despite that the cosmic evolution is defined by the Hubble parameter
as the rate of expansion and by the deceleration parameter
accelerated or decelerated expansion is determined, these two
parameters cannot clearly distinguish diverse models of DE. So, the
progression in cosmological data during the last decade leads us to
go beyond these two parameters. In order to perform more exact
calculation concerning this issue a new diagnostic pair for dark
energy has been proposed \cite{16}. This pair $(r,s)$ which is
called Statefinder is a geometrical diagnostic and allows us to
characterize the properties of dark energy. Its definition is as
follows

\begin{equation}\label{AF}
r=\frac{\dddot{a}}{aH^3}=1+\frac{\ddot{H}}{H^3}+3\frac{\dot{H}}{H^2},\hspace{5mm}s=\frac{r-1}{3(q-\frac{1}{2})}.
\end{equation}
To evaluate this pair in our model, all we need is to compute
$\frac{\ddot{H}}{H^3}$ which may be obtained by taking a derivative
of the relation (\ref{U}). The result is

\begin{equation}\label{AG}
\begin{split}
\frac{\ddot{H}}{H^3}=\frac{1}{1-\frac{\beta^2a^2}{(\alpha+\beta
a)^2\ln(\alpha+\beta a)^2}+\frac{2\beta a}{(\alpha+\beta
a)\ln(\alpha+\beta
a)}-\frac{1}{2}(c^2+\Omega_D)}\Big(\frac{\Omega'_D\beta
a}{(\alpha+\beta a)\ln(\alpha+\beta
a)}~~~~~~~~~~~~~~~~\\~~~~+\frac{3\beta^4a^4}{(\alpha+\beta
a)^4\ln(\alpha+\beta a)^2}(3+\frac{2}{\ln(\alpha+\beta
a)})+\frac{\beta^2a^2}{(\alpha+\beta a)^2\ln(\alpha+\beta
a)}(4-\Omega_D)~~~~~~~~~~~~~~\\~-\frac{2\beta^3a^3}{(\alpha+\beta
a)^3\ln(\alpha+\beta a)}(1+\frac{2}{\ln(\alpha+\beta
a)})-\frac{2\beta a}{(\alpha+\beta a)\ln(\alpha+\beta
a)}(\Omega_D-2)~~~~~~~~~~~~~~~
~~\\~~~+\frac{\beta^2a^2}{(\alpha+\beta a)^2\ln(\alpha+\beta
a)^2}(2-\Omega_D)+\frac{3}{2}(b^2+1)\Omega'_D-\big(\frac{\Omega_D\beta
a}{(\alpha+\beta a)\ln(\alpha+\beta
a)}(\Omega_D-2)~~~~~\\~~~+\frac{\beta^2a^2}{(\alpha+\beta
a)^2\ln(\alpha+\beta a)}(1-\frac{\beta a}{(\alpha+\beta
a)\ln(\alpha+\beta
a)})+\frac{3}{2}((b^2+1)\Omega_D-1)\big)~~~~~~~~~~~~~~~~~~\\~~~\big(\frac{2\beta^3a^3}{(\alpha+\beta
a)^3\ln(\alpha+\beta a)^2}(1+\frac{1}{\ln(\alpha+\beta
a)})+\frac{2\beta a}{(\alpha+\beta a)\ln(\alpha+\beta
a)}~~~~~~~~~~~~~~~~~~~~~~~~~~~~~\\~~~~-\frac{\beta^2a^2}{(\alpha+\beta
a)^2\ln(\alpha+\beta a)}(2+\frac{4}{\ln(\alpha+\beta
a)})-\frac{1}{2}\Omega'_D\big)\Big)+(\frac{\dot{H}}{H^2})^2.~~~~~~~~~~~~~~~~~~~~~~~~~~~~~~~~~
\end{split}
\end{equation}

\begin{figure}[t]
\begin{center}
\includegraphics[width=0.45\textwidth]{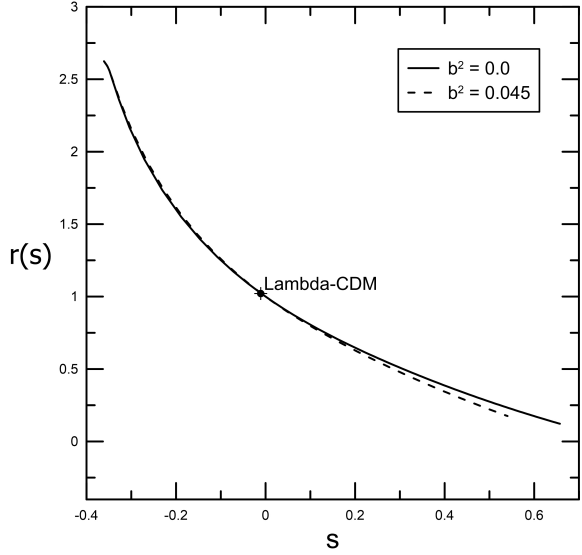}
\caption{Statefinder pair.} \label{fig s}
\end{center}
\end{figure}
With the help of this relation, we have plotted the statefinder pair
($r$ in terms of $s$) in figure \ref{fig s}. It is seen from this
figure that as the universe expands, by increasing the parameter $r$
the parameter $s$ decreases (from positive to negative values). The
point $(r,s)=(1,0)$ corresponds to the $\Lambda$CDM model. The state
finder trajectory indicates the Chaplygin gas behavior (where $s <
0, r > 1$) and the phantom like behavior (where $s > 0, r < 1$).

\section{Om-Diagnostic}
\begin{figure}[t]
\begin{center}
\includegraphics[width=0.45\textwidth]{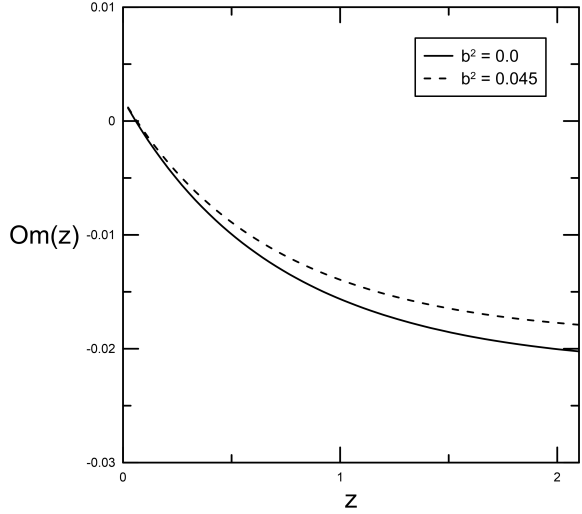}
\caption{The Om-diagnostic tool versus $z$.} \label{fig5}
\end{center}
\end{figure}

In order to study and differentiating different stages of the
universe, the Om-diagnostic tool has been proposed \cite{18}. Using
this tool and according to resulted curves in the final plot, one
can distinguish the behavior of DE model and divide it into two
sections. Phantom-like ($\omega_D<-1$) behavior corresponds to the
positive values in the Om(x) trajectories and quintessence
($\omega_D>-1$) comes from its negative value. The Om-diagnostic
tool can be explained as
\begin{equation}\label{DT}
\mbox{Om}(x)=\frac{h(x)^2-1}{x^3-1},
\end{equation}
where $h(x)=H(x)/H_0$ and $x=ln(z+1)^{-1}$. The evolution of
Om-diagnostic tool in terms of the redshift is plotted in figure
\ref{fig5}. It can be seen that for both non-interacting and
interacting models the trajectories present negative values which
show the quintessence behavior of the universe. Also, it is clear
that in the late time ($z<0$) the positive value of trajectories
implies phantom behavior which indicates uniformity with the
equation of state parameter as seen in figure \ref{fig2}.
\section{Data analysis methods}
In this section, we fit present model using the recent observational
data sets including SN Ia,
BAO and CMB. We have used the minimized chi-square test and obtain the best fit values of the free parameters for $1\sigma$ and $2\sigma$ confidence region.\\
For the SN Ia data sets, we use the cJLA data set of 31 check points
(30 bins) with the range of redshift $0.01<z<1.3$ \cite{19}. The
corresponding $\chi^2$ function is
\begin{equation}\label{SN}
\chi^2_{SNIa}=r^{t} C^{-1}_b r,
\end{equation}
in which
\begin{equation}\label{rSNIa}
r=\mu_b-M-5log_{10}d_L,
\end{equation}
where where $\mu_b$ is the observational distance modulus, $M$ is a
free normalization parameter and $C_b$ is the covariance matrix of
$\mu_b$, see Table F.2 in \cite{19}. Also, the dimensionless
luminosity distance is defined as

\begin{equation}\label{dL}
d_L=\frac{c(1+z)}{H_0}\int_{0}^{z'}
\frac{dz'}{[(1+z')^2(1+\Omega_mz')-z'(2+z')\Omega_D]^{0.5}}.
\end{equation}
For BAO we use data of BOSS DR12 including six BAO data points
\cite{20}. The $\chi^2_{BAO}$ function is defined as
\begin{equation}\label{BAO}
\chi^2_{BAO}=X^tC_{BAO}^{-1}X,
\end{equation}
which for $X$ we can write
\begin{equation}\label{XBAO}
X=\left(\begin{array}{c} \frac{D_M(0.38)r_{s,fid}}{r_s(z_d)}-1512.39\\
\frac{H(0.38)r_s(z_d)}{r_s(z_d)}-81.208\\
\frac{D_M(0.51)r_{s,fid}}{r_s(z_d)}-1975.22\\
\frac{H(0.51)r_s(z_d)}{r_s(z_d)}-90.9\\
\frac{D_M(0.61)r_{s,fid}}{r_s(z_d)}-2306.68\\
\frac{H(0.51)r_s(z_d)}{r_s(z_d)}-98.964\end{array}\right),
\end{equation}
where $r_{s,fid}=$147.78 Mpc is the sound horizon of fiducial model,
$D_M(z)=(1+z)D_A(z)$ is the comoving angular diameter distance. The
sound horizon at drag epoch $r_s(z_d)$ may be expressed as

\begin{equation}\label{BAO}
r_s(z_d)=\int_{z_d}^{\infty} \frac{c_s(z)}{H(z)}dz,
\end{equation}
where $c_s=1/\sqrt{3(1+R_b/(1+z))}$ is the sound speed with
$R_b=31500\Omega_bh^2(2.726/2.7)^{-4}$. The covariance matrix can be
downloaded from the online files of \cite{20}:
\begin{equation}\label{iCOVBAO}
C^{-1}_{BAO}=\begin{pmatrix}   624.707& 23.729  &325.332    &8.34963&   157.386 &3.57778\\
23.729  &5.60873    &11.6429    &2.33996    &6.39263    &0.968056\\
325.332 &11.6429    &905.777    &29.3392    &515.271&   14.1013\\
8.34963 &2.33996    &29.3392    &5.42327    &16.1422&   2.85334\\
157.386 &6.39263    &515.271    &16.1422    &1375.12&   40.4327\\
3.57778 &0.968056   &14.1013    &2.85334    &40.4327
&6.25936\end{pmatrix}.
\end{equation}
Probing the whole expansion history until the last scattering phase,
we use Planck 2015 for CMB \cite{21}. The $\chi^2_{CMB}$ function is
\begin{equation}\label{CMB}
\chi^2_{CMB}=q_i-q^{data}_i Cov^{-1}_{CMB}(q_i,q_j),
\end{equation}
where $q_1=R(z_*)$, $q_2=l_a(z_*)$ and $q_3=\omega_b$. $Cov_{CMB}$
is the covariance matrix \cite{21}. The data of Planck 2015 are
\begin{equation}\label{PLANCKDATA}
q^{data}_1=1.7382,\hspace{5mm} q^{data}_2=301.63,\hspace{5mm}
q^{data}_3=0.02262.
\end{equation}
The acoustic scale $l_A$ can be defined as
\begin{equation}\label{lA}
l_A=\frac{3.14d_L(z_*)}{(1+z)r_s(z_*)},
\end{equation}
where $r_s(z_*)$ is the comoving sound horizon at the decoupling
time ($z_*$). The redshift of the decoupling time is \cite{22}

\begin{equation}\label{z_*}
z_*=1048\left[1+0.00124(\Omega_bh^2)^{-0.738}\right]\left[1+g_1(\Omega_mh^2)^{g_2}\right],
\end{equation}
where
\begin{equation}\label{g1 g2}
g_1=\frac{0.0783(\Omega_bh^2)^{-0.238}}{1+39.5(\Omega_bh^2)^{-0.763}}, ~~~g_2=\frac{0.560}{1+21.1(\Omega_bh^2)^{1.81}}.
\end{equation}
The CMB shift parameter is \cite{23}
\begin{equation}\label{R}
R=\sqrt{\Omega_{m_0}}\frac{H_0}{c}r_s(z_*).
\end{equation}
Finally, the total $\chi^2$ is
\begin{equation}\label{TOTALX}
\chi^2_{total}=\chi^2_{SNIa}+\chi^2_{BAO}+\chi^2_{CMB}.
\end{equation}
By minimizing the above quantity we can perform the best-fit values
of the free parameters. Considering the $1\sigma$ and
$2\sigma$confidence level, the best-fit values of $\alpha$, $\beta$,
$c$ and $\Omega_{m0}$ are shown in table 1.

\begin{figure}[t]
\begin{center}
\includegraphics[width=0.75\textwidth]{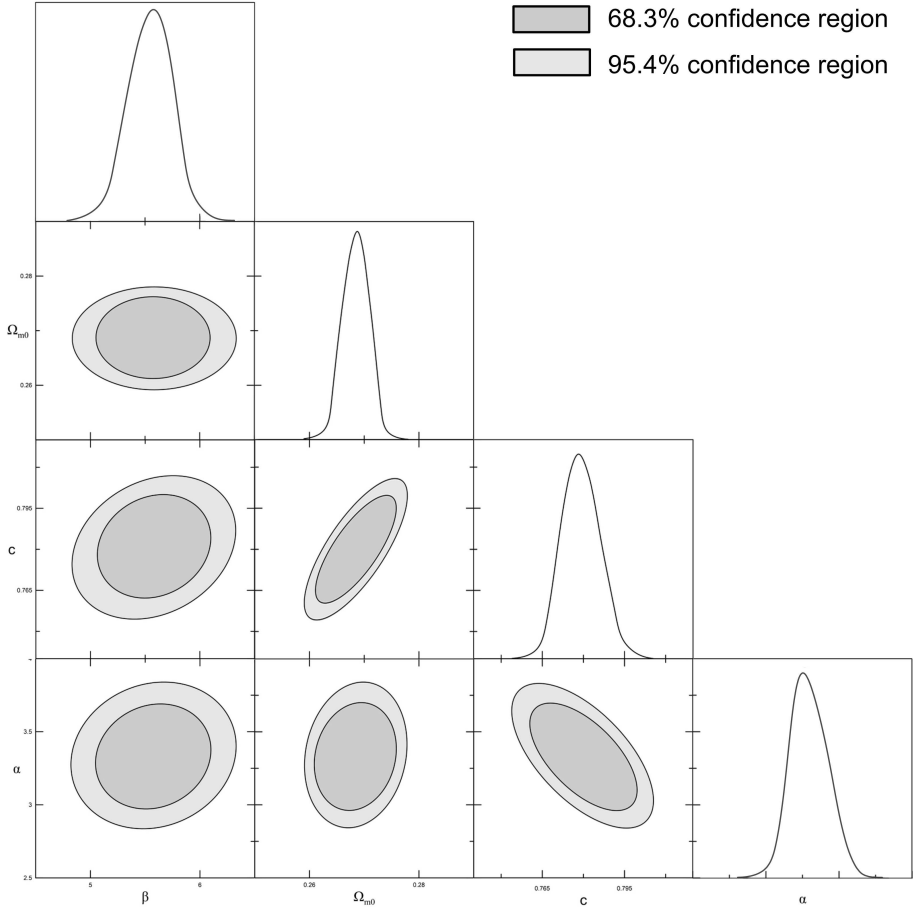}
\caption{The 2D 68.3\% and 95.4\% confidence level for $\alpha$,
$\beta$, $\Omega_{m0}$ and c.} \label{fig7}
\end{center}
\end{figure}

\begin{table}[h]
\begin{center}
\begin{tabular}{|c|c|c|}
\hline Parameters & cJLA + BOSS DR12 + Planck2015  \\ \hline
$\Omega_m$      & $0.268^{+0.008~+0.010}_{-0.007~-0.009}$
\\ \hline
$ \alpha$         & $3.361^{+0.332~+0.483}_{-0.401~-0.522}$
\\ \hline
$ \beta$          & $5.560^{+0.541~+0.780}_{-0.510~-0.729}$
\\ \hline
$c$                & $0.777^{+0.023~+0.029}_{-0.017~-0.023}$
\\ \hline
$b^2 $             & $0.045$          \\ \hline
$M $             & $10.7$ \\
\hline
\end{tabular}
\caption{68.3\% and 95.4\% error marginalized result for each
parameter.}
\end{center}
\end{table}
By using of the latest observational data sets namely cJLA, Boss
DR12 and Planck 2015, we have plotted 1D marginalized posterior
distributions and 2D confidence region of the important parameters
of the current model in figure \ref{fig7}.
\section{Summary}
In this paper we have applied a new HDE model (firstly proposed in
\cite{6} inspired by the DGP braneworld theory) to the BD gravity in
both interacting ($b^2=0.045$-fitted with the recent observational
data) and non-interacting ($b^2=0$) cases. The Hubble radius
$L=H^{-1}$ plays the role of the system's IR cutoff. Following
\cite{5}, we have taken a logarithmic BD scalar field which gives a
dynamical DP. Then we evaluated and plotted the density parameter,
EoS and DP in terms of redshift. We found that in both interacting
and non-interacting levels, the corresponding universe is expanding
with accelerating rate. Our numerical results shows that the
acceleration of accelerated expansion of the universe increases
exponentially after $z=0.75$. The DP for both interacting and
non-interacting models resulted an universe with accelerating
expansion and undergoes from matter to DE dominated universe
approximately at $z\approx 0.55$ which is consistent with recent
observational data $0.4<z<1$ \cite{17} and mimics the phantom dark
energy at the late time. We also performed the statefinder diagnosis
pair tool with different value of $c$ parameter which leads to
different trajectories in $s-r$ plane. Since the $\Lambda$CDM is the
main standard model of dark energy, hence we have tried to measure
the deviation of the points in $s-r$ plane from the point
$(r,s)=(1,0)$ correspond to the $\Lambda$CDM. Using the state finder
pair tool indicated a behavior similar to Chaplygin gas ($s<0,
r>1$).\\ In the language of the Om-diagnostic tool, its variation in
terms of the redshift by taking $x = ln(1 + z)^{-1}$, shows negative
values which implies the quintessence behavior. Finally, in order to
check compatibility with observational data and fitting the free
parameters, we used cJLA compilation for SNIa, six observational
points of BAO from BOSS DR12 and Planck 2015 for CMB. This
combination of the resent observational data sets results in
$\Omega_m=0.268^{+0.008~+0.010}_{-0.007~-0.009}$, $
\alpha=3.361^{+0.332~+0.483}_{-0.401~-0.522}$,
$\beta=5.560^{+0.541~+0.780}_{-0.510~-0.729}$,
$c=0.777^{+0.023~+0.029}_{-0.017~-0.023}$ and $b^2 =0.045$ with
$1\sigma$ and $2\sigma$ confidence interval.\vspace{5mm}\newline
\noindent {\bf Acknowledgement}\vspace{2mm}\noindent\newline We
would like to thank the referee for insightful comments which
improved the quality of the paper. This work has been supported
financially by Research council of the Central Tehran Branch,
Islamic Azad University.

\end{document}